\documentclass[%
reprint,
superscriptaddress,
showpacs,
amsmath,amssymb,
aps,
prf,
onecolumn,
longbibliography,
]{revtex4-2}

\usepackage{psfrag,graphicx,epsfig,color}
\usepackage{dcolumn}
\usepackage{bm}
\usepackage{natbib}
\usepackage{float}
\usepackage[usenames,dvipsnames,svgnames,table]{xcolor}
\usepackage{subfigure}
\usepackage{nicefrac}
\usepackage{listings}
\usepackage{xcolor}
\usepackage{footnote}

\newcommand{\ee}{\mathbf{e}}

\newcommand{\Rot}{\mathbf{U}}
\newcommand{\DD}{\mathbf{D}}
\newcommand{\SSs}{\mathbf{S}}
\newcommand{\ct}{\cos\theta}
\newcommand{\st}{\sin\theta}
\newcommand{\cpp}{\cos\phi}
\newcommand{\spp}{\sin\phi}

\newcommand{\TT}{\mathcal{T}}
\newcommand{\RR}{\mathcal{R}}
\newcommand{\re}{R_\lambda}

\begin{document}

\title{Relation between the moments of longitudinal velocity derivatives and of dissipation in turbulence }

\author{Ping-Fan Yang }
\affiliation{Institute of Extreme Mechanics, School of Aeronautics, National Key Laboratory of Aircraft Configuration Design, Key Laboratory for Extreme Mechanics of Aircraft of Ministry of Industry and Information Technology, Northwestern Polytechnical University, Xi’an 710072, Shaanxi, PR China}
\author{Haitao Xu}
\affiliation{Center for Combustion Energy, School of Aerospace Engineering and State Key Laboratory of Advanced Space Propulsion, Tsinghua University, Beijing 100084, PR China}
\author{Alain Pumir}
\affiliation{Ecole Normale Superieure, Universite de Lyon 1 and CNRS, 69007 Lyon, France}
\affiliation{French American Center for Theoretical Science, CNRS, KITP, University of California, Santa Barbara, California 93106-4030, USA}

\date{\today}

\begin{abstract}

In homogeneous and isotropic turbulence, measurements of the
longitudinal velocity derivative, $\partial_1 u_1$, make it possible to estimate
a surrogate of the rate of energy dissipation per unit mass, $\epsilon$: 
$\epsilon_s = 15 \nu (\partial_1 u_1)^2 $, where $\nu$ is the fluid viscosity,
in the sense that the averages of $\epsilon$ and $\epsilon_s$ are equal.
We show here that the
$n^{th}$ moments of the fluctuations $\epsilon$ and $\epsilon_s$, for
$n > 2$, are not exactly proportional to each other, and that the expression
for the moment $\langle \epsilon_s^n \rangle$ for $ n \ge 3$ involves in
addition to a term proportional to $\langle \epsilon^n \rangle$, other
contributions involving the invariant of the strain tensor, $\SSs$:
${\rm tr}( \SSs^3)$. The contribution of this term depends on the distribution 
of the dimensionless ratio 
$\mathcal{R} \equiv {\rm tr}(\SSs^3)/{\rm tr}(\SSs^2)^{3/2}$. We find, 
however, that the relation obtained by assuming that $\mathcal{R}$ is
uniformly distributed in the interval
$-1/\sqrt{6} \le \mathcal{R} \le 1/\sqrt{6}$, which is obtained when 
the matrix $\SSs$ has a Gaussian distribution, 
differs by no more than a few percents from the exact distribution.

\end{abstract}

\maketitle

\section{Introduction}

Velocity derivatives, $\partial_i u_j$, are essential quantities to 
describe the properties of turbulent flows. In particular, they are crucial
to understand dynamical aspects, including
fluid deformation, vorticity-strain interaction, stretching and amplification 
of scalar gradients. Moreover, velocity derivatives are related to the 
fundamental parameter characterising turbulence, the energy dissipation rate 
$\epsilon = 2 \nu S_{ij} S_{ij}$, where 
$S_{ij} \equiv 1/2 (\partial_i u_j + \partial_j u_i)$ is the rate of 
strain. Therefore, the statistical properties of velocity derivatives have 
been investigated extensively~\cite{frisch:95,sreeni:97,Johnson:24}. In fact, 
the very 
strong fluctuations of the rate of energy dissipation in turbulent flows,
which 
are at the root of the phenomenon of intermittency, leading to
deviations from the 
famous Kolmogorov theory~\cite{K41a}, were first revealed by studying the 
fourth moment of the longitudinal velocity derivative $\partial_1 u_1$ from 
hot-wire measurements in a wind tunnel~\cite{batchelor:49}. 
For historical and practical reasons, the quantity 
$\epsilon_s \equiv 15 \nu (\partial_1 u_1)^2 $ defined with the longitudinal 
velocity derivate only has been widely used as a surrogate of 
the energy dissipation rate $\epsilon$ because $S_{11} = \partial_1 u_1$ 
can be obtained 
using a single hot-wire probe in intense turbulent flows with large mean velocities such as wind tunnels and atmospheric boundary layers, and because their averages $\langle \epsilon_s \rangle$ and $\langle \epsilon\rangle$ are equal in homogeneous and isotropic turbulence (HIT). 
In the case $n=2$, an exact relation between $\langle \epsilon^2 \rangle$ 
and $\langle \epsilon_s^2 \rangle$
establishing a proportionality between
$\langle (\partial_1 u_1)^4 \rangle$ and $\langle (S_{ij}S_{ji})^2 \rangle$
 has been obtained by Siggia~\cite{siggia:81}.
This is an important information, since an outstanding question in 
turbulence theory is to characterize
the nature of the fluctuations of $\epsilon$, or equivalently, the moments 
of the distribution. 
Whether the moments of order 
$n \geq 3$ of $\epsilon$ and $\epsilon_s$  are
simply related to each other is the question asked in this note.

Boschung~\cite{boschung:15}, using the generating function formalism,
derived a general expression for the ratio 
of $\langle \epsilon^n \rangle / \langle \epsilon_s^n \rangle$ for all integers
$n$, which agree with the known values when $n=1$ and $n=2$. In this short 
article, we show that the general relation in \cite{boschung:15} is actually 
not exact.

As the longitudinal velocity derivative $\partial_1 u_1$ is equal to $S_{11}$, one of the components of the strain tensor,
one expects that the $n^{th}$ moments of $\epsilon_s$, or equivalently the 
$2n^{th}$ moment of $S_{11}$, are expressible as functions of the two non-zero 
invariants
of $\SSs$, namely ${\rm tr} (\SSs^2)$ and ${\rm tr} (\SSs^3)$~\cite{itskov:13}. 
We will denote throughout this text $\TT_n = {\rm tr}( \SSs^n ) $.
Elementary arguments
of homogeneity then suggest that $\langle (S_{11})^{2n}\rangle$ or $\langle \epsilon_s^n \rangle/(15 \nu)^n$ can be
expressed as sum of terms of the form $\langle \TT_2^m \TT_3^{2p} \rangle$,
such that $m + 3p = n$:
\begin{align}
\langle (S_{11})^{2n} \rangle = \sum_{2m + 6p = 2n} A^{(n)}_{m,p} \langle \TT_2^m \TT_3^{2p}  \rangle \, .
\label{eq:mom_general}
\end{align}
Here, we show how to determine the coefficients $A^{(n)}_{m,p}$, and in fact
provide a simple exact expression for $A^{(n)}_{n,0}$.
We further observe that for $0 \le m \le 2$, when the ratio between
$\langle \TT_2^m \TT_3^{2} \rangle/\langle \TT_2^{m + 3} \rangle = 1/18$, Eq.~\eqref{eq:mom_general} coincide with
the expressions given in~\cite{boschung:15} for $6 \le n \le 10$.
We show that such an identity is always satisfied provided the distribution 
of the dimensionless ratio, $\mathcal{R} \equiv \TT_3/\TT_2^{3/2}$ is uniformly
distributed in the interval $-1/\sqrt{6} \le \mathcal{R} \le 1/\sqrt{6}$,
as it happens for a Gaussian ensemble
of traceless random matrices. Such a distribution, however, does not hold for
turbulent flows, because of the well known condition that
$\langle \TT_3 \rangle < 0$~\cite{betchov:56}. Consistent with the results
of~\cite{BPB22}, we show that the distribution of the ratio $\mathcal{R}$, 
conditioned on $\TT_2$, is close to uniform for
$\TT_2 \lesssim 10^{-2} \langle \TT_2 \rangle$, but deviates significantly
for larger values of $\TT_2$, which leads to a small difference in the
ratio $\langle \TT_2^m \TT_3^{2} \rangle/\langle \TT_2^{m + 3} \rangle $.
The resulting contribution to Eq.~\eqref{eq:mom_general}, however, remain
numerically very small compared to the expressions
in~\cite{boschung:15}, which therefore provide a good approximation of 
the moments of $\epsilon$, accurate to a few percents.

\section{Expression of the even moments of the longitudinal velocity derivative}
\label{sec:derivation}

We consider in this work homogeneous and isotropic turbulent flows. The 
relations we obtain here does not require the flow to be statistically steady. 
For unsteady flows, such as decaying turbulence, the results derived
in this section can be used to relate the moments of $\epsilon$ and 
$\epsilon_s$ at any instant.
As the rate of strain tensor $\SSs$ 
is given by a symmetric matrix, it can be diagonalized in an orthogonal frame, 
i.e., $\SSs = \Rot \DD \Rot^T$,
where $\Rot$ is a rotation matrix, satisfying $\Rot  \Rot^T = \mathbf{I_d}$,
and $\DD$ is a diagonal matrix, $\DD = {\rm diag}( \lambda_1, \lambda_2 ,
\lambda_3)$, where $\lambda_i$ are the eigenvalues of $\mathbf{S}$, satisfying
$\lambda_1 + \lambda_2 + \lambda_3 = 0$ (by incompressibility).
We denote the eigenvectors of $\SSs$, associated with the eigenvalues
$\lambda_i$, as $\ee_i$.
Any unit vector $\hat{\mathbf{x}}$ can then be decomposed in the frame formed by the eigenvectors of $\SSs$ as
$\hat{\mathbf{x}} = c_1 \ee_1 + c_2 \ee_2 + c_3 \ee_3$. 
The coordinates $(c_1, c_2 , c_3 )$ parametrize a point
on the unit sphere. A convenient parametrization is given by:
\begin{align}
c_1 = \ct  ~~, ~~~~ c_2 = \st \cpp ~~ {\rm and} ~~~~ c_3 = \st \spp  \ .
\label{eq:c_i}
\end{align}
The longitudinal velocity derivative along the direction of $\hat{\mathbf{x}}$, which is equivalent to the component $S_{11}$ of $\SSs$, can be expressed
as a function of the eigenvalues, $\lambda_i$, and the coordinates $c_i$ 
by writing:
\begin{align}
S_{11} = \hat{\mathbf{x}} \cdot \SSs \cdot \hat{\mathbf{x}} =  c_1^2 \lambda_1 + c_2^2 \lambda_2 + c_3^3 \lambda_3 \, .
\label{eq:S_vs_lam}
\end{align}

The derivation of the explicit expressions rests on a few identities, which
we review in turn.

\paragraph*{The multinomial theorem.}
Being interested in the $2n^{th}$ moment of $S_{11}$, we need to express 
$(c_1^2 \lambda_1 + c_2^2 \lambda_2 + c_3^2 \lambda_3)^{2n}$. This can be done using the so-called
multinomial theorem: 
\begin{align}
(z_1 + z_2 +  z_3)^{2n} = \sum_{k_1 + k_2 + k_3 = 2 n} { 2 n \choose {k_1 ~ k_2 ~ k_3} } z_1^{k_1} z_2^{k_2} z_3^{k_3} ~ ,
\label{eq:multinomial}
\end{align}
where:
\begin{align}
{ n \choose{ k_1 ~ k_2 ~ k_3} } = \frac{n!}{k_1! k_2! k_3!} \, .
\label{eq:def_coeff}
\end{align}

\paragraph*{Angular averages.}
The averages over the direction cosines $c_i$ can be computed explicitly by
integrating over the unit sphere. For a unit vector
$\mathbf{c}=(c_1,c_2,c_3)$ uniformly distributed on the sphere in three
dimensions, we write
\begin{align}
\left\langle c_1^{2a}c_2^{2b}c_3^{2c}\right\rangle
= \frac{1}{4\pi}\int_{S^2} c_1^{2a}c_2^{2b}c_3^{2c}\, d\Omega ,
\end{align}
with $d\Omega$ the solid angle element. Using spherical coordinates,
\begin{align}
c_1 = \cos\theta,\qquad
c_2 = \sin\theta\cos\phi,\qquad
c_3 = \sin\theta\sin\phi ,
\end{align}
and $d\Omega = \sin\theta\, d\theta\, d\phi$ with
$0\le\theta\le\pi$ and $0\le\phi\le 2\pi$, we obtain
\begin{align}
\left\langle c_1^{2a}c_2^{2b}c_3^{2c}\right\rangle
&= \frac{1}{4\pi}
\int_0^{2\pi}\!\!\int_0^\pi
(\cos\theta)^{2a}\,(\sin\theta\cos\phi)^{2b}\,
(\sin\theta\sin\phi)^{2c}\,
\sin\theta\, d\theta\, d\phi \nonumber\\
&= \frac{1}{4\pi}
\left[\int_0^\pi \sin^{2(b+c)+1}\theta\,\cos^{2a}\theta\, d\theta\right]
\left[\int_0^{2\pi}\cos^{2b}\phi\,\sin^{2c}\phi\, d\phi\right] \nonumber\\
&= \frac{1}{4\pi}\, I_\theta\, I_\phi .
\end{align}
By symmetry of the integrands, these integrals can be reduced to integrals
over $[0,\pi/2]$ and expressed in terms of the Beta function. Using
\begin{align}
\int_0^{\pi/2}\sin^{2\alpha-1}x\,\cos^{2\beta-1}x\, dx
= \frac{1}{2}B(\alpha,\beta)
= \frac{1}{2}\frac{\Gamma(\alpha)\Gamma(\beta)}{\Gamma(\alpha+\beta)} ,
\label{eq:def_beta}
\end{align}
we find
\begin{align}
I_\phi &= 4\int_0^{\pi/2}\cos^{2b}\phi\,\sin^{2c}\phi\, d\phi
= 2\,\frac{\Gamma\!\left(b+\tfrac12\right)
          \Gamma\!\left(c+\tfrac12\right)}
         {\Gamma(b+c+1)} \, , \label{eq:I_phi} \\
I_\theta &= 2\int_0^{\pi/2}
\sin^{2(b+c)+1}\theta\,\cos^{2a}\theta\, d\theta
= \frac{\Gamma(b+c+1)\Gamma\!\left(a+\tfrac12\right)}
       {\Gamma\!\left(a+b+c+\tfrac32\right)} \, . \label{eq:I_theta}
\end{align}
Combining these results,
\begin{align}
\left\langle c_1^{2a}c_2^{2b}c_3^{2c}\right\rangle
= \frac{1}{2\pi}\,
\frac{\Gamma\!\left(a+\tfrac12\right)
      \Gamma\!\left(b+\tfrac12\right)
      \Gamma\!\left(c+\tfrac12\right)}
     {\Gamma\!\left(a+b+c+\tfrac32\right)} .
\label{eq:ang_gamma}
\end{align}
For half-integer arguments, the Gamma function can be expressed in terms of
double factorials:
\begin{align}
\Gamma\!\left(n+\tfrac12\right)
= \frac{(2n-1)!!}{2^n}\sqrt{\pi}\qquad (n=0,1,2,\ldots) .
\end{align}
Using this identity in Eq.~\eqref{eq:ang_gamma}, one obtains after
straightforward algebra: 
\begin{align}
\left\langle c_1^{2a}c_2^{2b}c_3^{2c}\right\rangle
= \frac{(2a-1)!!(2b-1)!!(2c-1)!!}{\big(2(a+b+c)+1\big)!!} ,
\label{eq:ang_general}
\end{align}
where $(-1)!!=1$ and $(2n+1)!!=(2n+1)(2n-1)\cdots 3\cdot 1$.
Equation~\eqref{eq:ang_general} is the formula used to compute all angular
averages appearing in the expression of the sixth and eighth moments.

\subsection{Determination of the coefficients $A^{(n)}_{n,0}$ in Eq.~\eqref{eq:mom_general}}
\label{subsec:A^n_n_0}

The general form given by Eq.~\eqref{eq:mom_general} can be derived by starting
from the general expression for $S_{11}$, Eq.~\eqref{eq:S_vs_lam}, and expanding
systematically the various terms using Eq.~\eqref{eq:multinomial}, with the
explicit form of the angular averages~\eqref{eq:ang_general}. The more
efficient approach proposed here consists in postulating directly the expected
form~\eqref{eq:mom_general}, and establishing relations between the
coefficients $A^{(n)}_{m,p}$ (with $ 2 m + 6p = 2n$). We checked that both
approaches lead to the identical results.

We begin by the coefficient $A^{(n)}_{n,0}$, for which we can establish an elementary
expression. To this end, we chose the eigenvalues of $\SSs$ to be $\lambda_1 = 0$, $\lambda_2 = - \lambda_3 = 1$. In this case, the invariant $\TT_2 = 2$,
and $\TT_3 = 0$, so $\langle (\partial_x u)^{2n} \rangle = 2^n A^{(n)}_{n,0}$.
With the postulated form of the eigenvalues, Eq.~\eqref{eq:S_vs_lam} can be
expressed as: $S_{11} = \sin^2 \theta \cos 2\phi$, whose moment of order $2n$
is:
\begin{eqnarray}
\langle S_{11}^{2n} \rangle & = & \frac{1}{4 \pi} \left( \int_0^{\pi} \sin^{4n+1} \theta d\theta \right) \left( \int_0^{2 \pi} \cos^{2n} ( 2\phi) d \phi \right) \nonumber \\
& = & \left( \frac{1}{2} \int_0^{\pi} \sin^{4n+1} \theta d\theta  \right) \times
\left( \frac{1}{2 \pi} \int_0^{2 \pi}  \cos^{2n} ( 2\phi) d \phi \right) \, . \label{eq:S_11_step1}
\end{eqnarray}
The two integrals in Eq.~\eqref{eq:S_11_step1} can be exactly expressed, with the help of Eqs.~\eqref{eq:def_beta}. Specifically, the
integral in $\theta$ has the exact expression (see also ~\cite{GR}):
\begin{align}
\frac{1}{2} \int_0^{\pi} \sin^{4n+1} \theta d\theta  = \frac{4n!!}{(4n+1)!!} \, .
\label{eq:int1}
\end{align}
The second integral in Eq.~\eqref{eq:S_11_step1} can be expressed, using an
elementary change of variables as:
\begin{align}
\frac{1}{2 \pi} \int_0^{2 \pi}  \cos^{2n} ( 2\phi) d \phi = \frac{2}{\pi}
\int_0^{\pi/2} \cos^{2n} (u) du = \frac{(2n-1)!!}{2n!!} \, ,
\label{eq:int2}
\end{align}
the last equality can be found from the exact expression 
Eq.~\eqref{eq:def_beta}; see also
Ref.~\cite{GR}. Combining Eqs.~\eqref{eq:S_11_step1}, \eqref{eq:int1} and \eqref{eq:int2}, one otbains:
\begin{align}
A^{(n)}_{n,0} = \frac{1}{2^n} \langle S_{11}^{2n} \rangle = \frac{(2n)!}{n!} \frac{(2n-1)!!}{(4n+1)!!} = \frac{(n+1) (n+2) ... 2n}{(2n+1) (2n+3) ... (4n+1)}  \, ,
\label{eq:A_n_n}
\end{align}
so the numerator consists of a product over the $n$ consecutive integers,
from $n+1$ till $2n$, and the denominator of a product of the $n+1$ consecutive
odd integers, from $(2n+1)$ to $(4n + 1)$.
One can readily checks that this expression coincides with the known expressions
for $n = 1$ and $n = 2$:
\begin{align}
\langle (S_{11})^2 \rangle = \frac{2}{15} \langle \TT_2 \rangle ~~~~ {\rm and} ~~~~ \langle ( S_{11})^4 \rangle = \frac{4}{105} \langle \TT_2^2 \rangle \,.
\label{eq:check_A_n_n}
\end{align}
For $n \ge  3$, the explicit expressions for $A^{(n)}_{n,0}$ are:
\begin{align}
A^{(3)}_{3,0} = \frac{40}{3003} ~, ~~ A^{(4)}_{4,0} = \frac{112}{21879}, ~~ A^{(5)}_{5,0} = \frac{96}{46189}~ , ~~ {\rm and} ~~ A^{(6)}_{6,0} = \frac{2112}{2414425} \, .
\label{eq:A_3_0}
\end{align}

Using Stirling's formula, one finds that for $n \gg 1$,
$A^{(n)}_{n,0} \sim    2^{-(n+3/2)}/n$. 

\subsection{Determination of the coefficients $A^{(n)}_{n-3,1}$ in Eq.~\eqref{eq:mom_general}}
\label{subsec:A^n-3_n_1}

Determining $A^{(n)}_{n,0}$ is sufficient to completely characterize
the relation between the moment of order $n$ of $\TT_2$ and the moment
of order $2n$ of $\partial_1 u_1$ for $n \le 2$. However, for $ 3 \le n \le 5$,
one needs another coefficient, namely $A^{(n)}_{n-3,1}$, which corresponds
to a term $\langle \TT_2^{n-3} \TT_3^2 \rangle$ in Eq.~\eqref{eq:mom_general}.
For $n \ge 6$, one needs in addition to determine $A^{(n)}_{n-6,2}$, which
we will discuss in Section~\ref{subsec:A^n-p_n_p}.
The determination of the coefficients $A^{(n)}_{n-3,1}$ etc can be obtained
by considering other configurations of the eigenvalues of $\SSs$.

To this effect, we propose using the following:
 $(\lambda_1,\lambda_2,\lambda_3)=(2,-1,-1)$.

In this case,
\begin{align}
\TT_2 = 2^2+(-1)^2+(-1)^2 = 6 ~,~~~
\TT_3 = 2^3+(-1)^3+(-1)^3 = 6 \, ,
\end{align}
so that
\begin{align}
\TT_2^4 = 6^4 = 1296 ~~~{\rm and}~~~ \TT_3^2 \TT_2 = 6^2 \cdot 6 = 216 \, .
\end{align}
Equation~\eqref{eq:mom_general} now reads, for $ 3 \le n \le 5$:
\begin{align}
\langle ( S_{11})^{2n} \rangle = 6^n A^{(n)}_{n,0} + 6^{n-1} A^{(n)}_{n-3,1} \, .
\label{eq:B_case2_struct}
\end{align}
With this choice of eigenvalues,
\begin{align}
S_{11} = 2c_1^2 - c_2^2 - c_3^2 = 2c_1^2 - (1-c_1^2) = 3c_1^2 - 1 \, ,
\end{align}
so that
\begin{align}
(S_{11})^{2n} = (3c_1^2-1)^{2n}
= \sum_{p=0}^{2n} {2n \choose p} 3^p (-1)^{2n-p} c_1^{2p} \, .
\end{align}
Using again~\eqref{eq:ang_general} with $b=c=0$,
$\langle c_1^{2p}\rangle = (2p-1)!!/(2p+1)!!$, and summing over $p$,
one obtains
\begin{align}
\langle (S_{11})^{6} \rangle = \frac{3392}{1001} ~~~, ~~~
\langle (S_{11})^{8} \rangle = \frac{24320}{2431} ~~~ {\rm and}
\langle (S_{11})^{10} \rangle = \frac{1483776}{46189} \, .
\label{eq:B_case2_val}
\end{align}
Substituting Eq.~\eqref{eq:A_3_0} and Eq.~\eqref{eq:B_case2_val} into
Eq.~\eqref{eq:B_case2_struct} yields
\begin{align}
A^{(3)}_{0,1} = \frac{128}{9009} ~~~, ~~~
A^{(4)}_{1,1} = \frac{1024}{65637} ~~~ {\rm and}  ~~~
A^{(5)}_{2,1} = \frac{5120}{415701} \, .
\end{align}

Together with the expressions of $A^{(n)}_{n,0}$ determined in
Section~\ref{subsec:A^n_n_0}, Eq.~\eqref{eq:mom_general} completely determines
the momments of $S_{11}$ up to order $10$:
\begin{eqnarray}
\langle S_{11}^6 \rangle & = & \frac{40}{3003} \langle \TT_2^3 \rangle  + \frac{128}{9009} \langle \TT_3^2 \rangle  \, , \label{eq:6_mom} \\
\langle S_{11}^8 \rangle & = & \frac{112}{21879} \langle \TT_2^4 \rangle  + \frac{1024}{65637} \langle \TT_3^2 \TT_2 \rangle  \, , \label{eq:8_mom} \\
\langle S_{11}^{10} \rangle & = & \frac{96}{46189} \langle \TT_2^5 \rangle  + \frac{5120}{415701} \langle \TT_3^2 \TT_2^2 \rangle  \, . \label{eq:10_mom}
\end{eqnarray}
These expressions have been extensively checked by explicit calculations
(not shown here).

\subsection{Determination of the coefficients $A^{(n)}_{n-6p,p}$ in Eq.~\eqref{eq:mom_general}}
\label{subsec:A^n-p_n_p}

For $n \ge 6$, coefficients of the form $A^{(n)}_{\,n-6,2}$ appear in
Eq.~\eqref{eq:mom_general}.  
To determine these coefficients, one may generalize the strategy used in
Secs.~\ref{subsec:A^n_n_0} and \ref{subsec:A^n-3_n_1} by considering a family of
explicit eigenvalue triplets that span different combinations of the invariants
$\TT_2$ and $\TT_3$. A convenient choice is to take
\begin{equation}
(\lambda_1,\,\lambda_2,\,\lambda_3)
   = (i,\; i+1,\; -2i-1),
   \qquad i = 1,\;2,\;\ldots,\; \lfloor n/3 \rfloor + 1 ,
\end{equation}
where $\lfloor x \rfloor$ denotes the \emph{floor function}, i.e.\ the largest
integer not exceeding $x$.  
This parametrization automatically satisfies $\lambda_1+\lambda_2+\lambda_3=0$
and provides a sufficiently rich set of algebraic constraints to extract all
required coefficients $A^{(n)}_{\,n-6p,p}$.

Following the same computational spirit as in the lower-order cases, we evaluate
$\TT_2 = \lambda_1^2+\lambda_2^2+\lambda_3^2$ and 
$\TT_3 = \lambda_1^3+\lambda_2^3+\lambda_3^3$ for each admissible~$i$, expand
$S_{11}^{2n}$ using Eqs.~\eqref{eq:S_vs_lam} and~\eqref{eq:multinomial}, and solve the resulting linear
system for the unknown coefficients.  For completeness, we provide in
Appendix~\ref{app:mathematica} a \texttt{Mathematica} script implementing this
procedure for arbitrary~$n$. Applying this method for $n=6,7$ and $8$  leads to the explicit expression:
\begin{eqnarray}
\langle S_{11}^{12} \rangle &= 
 \frac{2112}{2414425} \langle \TT_2^6 \rangle 
 + \frac{4096}{482885} \langle \TT_3^2 \TT_2^3 \rangle 
 + \frac{32768}{39113685} \langle \TT_3^4 \rangle ,
\label{eq:12_mom} \\[1ex]
\langle S_{11}^{14} \rangle &= 
 \frac{18304}{48474225} \langle \TT_2^7 \rangle 
 + \frac{157696}{29084535} \langle \TT_3^2 \TT_2^4 \rangle
 + \frac{458752}{261760815} \langle \TT_3^4 \TT_2 \rangle ,
\label{eq:14_mom} \\[1ex]
\langle S_{11}^{16} \rangle &= 
 \frac{3328}{20036013} \langle \TT_2^8 \rangle 
 + \frac{2981888}{901620585} \langle \TT_3^2 \TT_2^5 \rangle
 + \frac{3670016}{1622917053} \langle \TT_3^4 \TT_2^2 \rangle .
\label{eq:16_mom}
\end{eqnarray}

\section{Comparison with earlier work}

 The existence of a relation between the moments of order $2n$ of
$\partial_1 u_1$, and the moments of order $n$ of $\TT_2$, beyond
the relations established for $n = 1$ and $n = 2$
has been postulated by Ref.~\cite{boschung:15}, who
used the generating function formalism to obtain: 
\begin{align}
\langle ( S_{11})^6 \rangle  =  \frac{8}{567} \langle \TT_2^3 \rangle ~ , ~~~
\langle ( S_{11})^8 \rangle  =  \frac{16}{2673} \langle \TT_2^4 \rangle  ~ , ~~~
\langle ( S_{11})^{10} \rangle  =  \frac{32}{11583} \langle \TT_2^5 \rangle ~~~ {\rm and} ~~~ \langle (S_{11})^{12} \rangle = \frac{64}{47385} \langle \TT_2^6 \rangle~ , \label{eq:mom_ps}
\end{align}
which do not involve the quantity $\TT_3 $ nor any of its moments.

We observe that the two sets of expressions (\ref{eq:6_mom}-\ref{eq:10_mom})
and \eqref{eq:mom_ps}) produce identical results when:
\begin{align}
\frac{\langle \TT_3^2  \rangle}{\langle \TT_2^3 \rangle } =
\frac{\langle \TT_3^2 \TT_2 \rangle }{ \langle \TT_2^4 \rangle } =
\frac{\langle \TT_3^2 \TT_2^2 \rangle }{\langle \TT_2^5  \rangle } = \frac{1}{18} \, .
\label{eq:ratio}
\end{align}
These relations can be understood by assuming a
particular distribution of the eigenvalues of $\SSs$, as we explain
in the following subsection. 

\subsection{ Statistical properties of the eigenvalues of ${\rm tr}(\SSs^3)/{\rm tr}(\SSs^2)^{3/2}$ }
\label{subsec:stat_eigenvalues}

 We focus here on the dimensionless ratio
$\mathcal{R} = \TT_3/\TT_2^{3/2}$:
\begin{align}
\mathcal{R} = \frac{\TT_3  }{ \TT_2^{3/2} } = \frac{(\lambda_1^3  + \lambda_2^3 + \lambda_3^3 ) }{( \lambda_1^2 + \lambda_2^2 + \lambda_3^2)^{3/2} }
= 3 \frac{\lambda_1 \lambda_2 \lambda_3 }{( \lambda_1^2 + \lambda_2^2 + \lambda_3^2)^{3/2} }  \, ,
\label{eq:def_rho}
\end{align}
and depends only on the relative distribution of the 3 eigenvalues of $\SSs$.
We assume for now that the eigenvalues of $\SSs$ are ranked in decreasing order:
$\lambda_1 \ge \lambda_2 \ge \lambda_3$. Unless the three eigenvalues are all
$0$, $\lambda_1 > 0$, and we define $\zeta = \lambda_2/\lambda_1$, which
satisfies $ -1/2 \le \zeta \le 1$. In terms of $\zeta$, $\mathcal{R}$ can be
expressed as:
\begin{align}
\mathcal{R} = - \frac{3}{2 \sqrt{2} } \frac{\zeta + \zeta^2}{(1 + \zeta + \zeta^2)^{3/2}}  \, .
\end{align}
Elementary algebra shows that $\mathcal{R}$ is a decreasing function of
$\zeta$, and that its values are in the interval
$-\frac{1}{\sqrt{6}} \le \mathcal{R} \le \frac{1}{\sqrt{6}}$.

Now, let's introduce the notation $P_2(\TT_2)$ and $P_3(\TT_3)$ as the probability distribution functions (PDFs) of $\TT_2$ and $\TT_3$,
and $P_{2,3}(\TT_2, \TT_3)$ as the joint PDF of the two.
This leads to the following expression for the moments $\langle \TT_3^{2} \TT_2^p \rangle$:
\begin{eqnarray}
\langle \TT_3^2 \TT_2^p \rangle & = & \int d \TT_2 d \TT_3 P_{2,3} (\TT_2, \TT_3) \TT_3^2 \TT_2^p \nonumber \\
& = & \int d \TT_2 P_2 (\TT_2) \TT_2^p \left[ \int d \TT_3 \frac{P_{2,3}(\TT_2, \TT_3)}{P_2(\TT_2)} \TT_3^2  \right] \nonumber \\
& = & \int d \TT_2 P_2 (\TT_2) \TT_2^{p+3} \left[ \int d \TT_3 \frac{P_{2,3}(\TT_2, \TT_3)}{P_2(\TT_2)} \frac{\TT_3^2}{\TT_2^{3}} \right]  \, .
\label{eq:T_3^2_T2}
\end{eqnarray}
The last expression in Eq.~\eqref{eq:T_3^2_T2} suggests using the
change of variable $\TT_3 = \mathcal{R} \times \TT_2^{3/2}$, which leads to:
\begin{align}
\langle \TT_3^2 \TT_2^p \rangle  =
\int d \TT_2 P_2 (\TT_2) \TT_2^{p+3} \left[ \int d \mathcal{R} P_c ( \mathcal{R} | \TT_2) \mathcal{R}^2 \right] \, ,
\label{eq:mom_p_2}
\end{align}
where
$P_c( \mathcal{R} | \TT_2) = P_{2,3}(\TT_2, \TT_2^{3/2} \mathcal{R} )/P_2(\TT_2)$ is the conditional PDF of $\mathcal{R}$ when $\TT_2$ is given.

Assuming that the distribution of $\mathcal{R}$ is uniform between $-1/\sqrt{6}$
and $1/\sqrt{6}$ and independent of $\TT_2$:
\begin{align}
P_c (\mathcal{R} | \TT_2) = \frac{\sqrt{6}}{2}~~~~ {\rm for}  -1/\sqrt{6} \le \mathcal{R} \le 1/\sqrt{6}
\label{eq:uniform_R}
\end{align}
allows  us to establish precisely the relations given
by Eq.~\eqref{eq:ratio}:
$\langle \TT_3^2 \TT_2^p \rangle = 1/18 \langle \TT_2^{p+3} \rangle$.

Whether the eigenvalues of $\SSs$ are such that $\mathcal{R}$
is uniformly distributed i.e. $P_c( \RR | \TT_2)$
satisfies Eq.~\eqref{eq:uniform_R}, is by no means correct~\cite{ashurst:87}.
In fact,
it is known empirically that the intermediate eigenvalue of $\SSs$ is
predominantly positive, so the distribution of $\RR$ cannot be uniform.
We also point here to the numerical study of the conditional value of
$\beta \equiv \sqrt{6} \lambda_2/(\lambda_1^2 + \lambda_2^2 + \lambda_3^2)^{1/2}$,
conditioned on $\TT_2$, which shows that
$\langle \beta | \TT_2 \rangle \approx 0$ for
$\TT_2 /\langle \TT^2 \rangle \ll 1$, but
$\langle \beta | \TT_2 \rangle \approx 0.5$ for
$\TT_2 /\langle \TT^2 \rangle \gg 1$~\cite{BPB22}

\begin{figure}[t]
\begin{center}
\subfigure[]{
\includegraphics[scale=0.36]{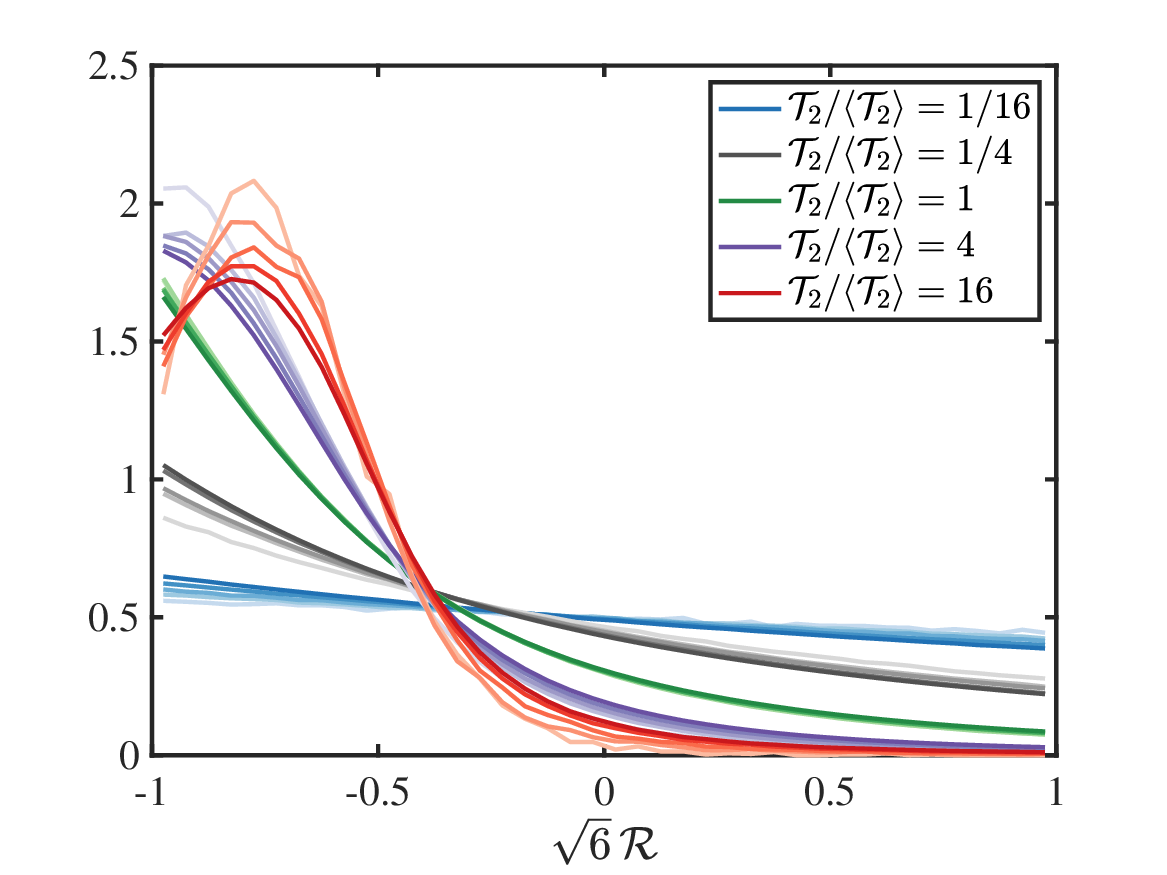}}
\subfigure[]{
\includegraphics[scale=0.36]{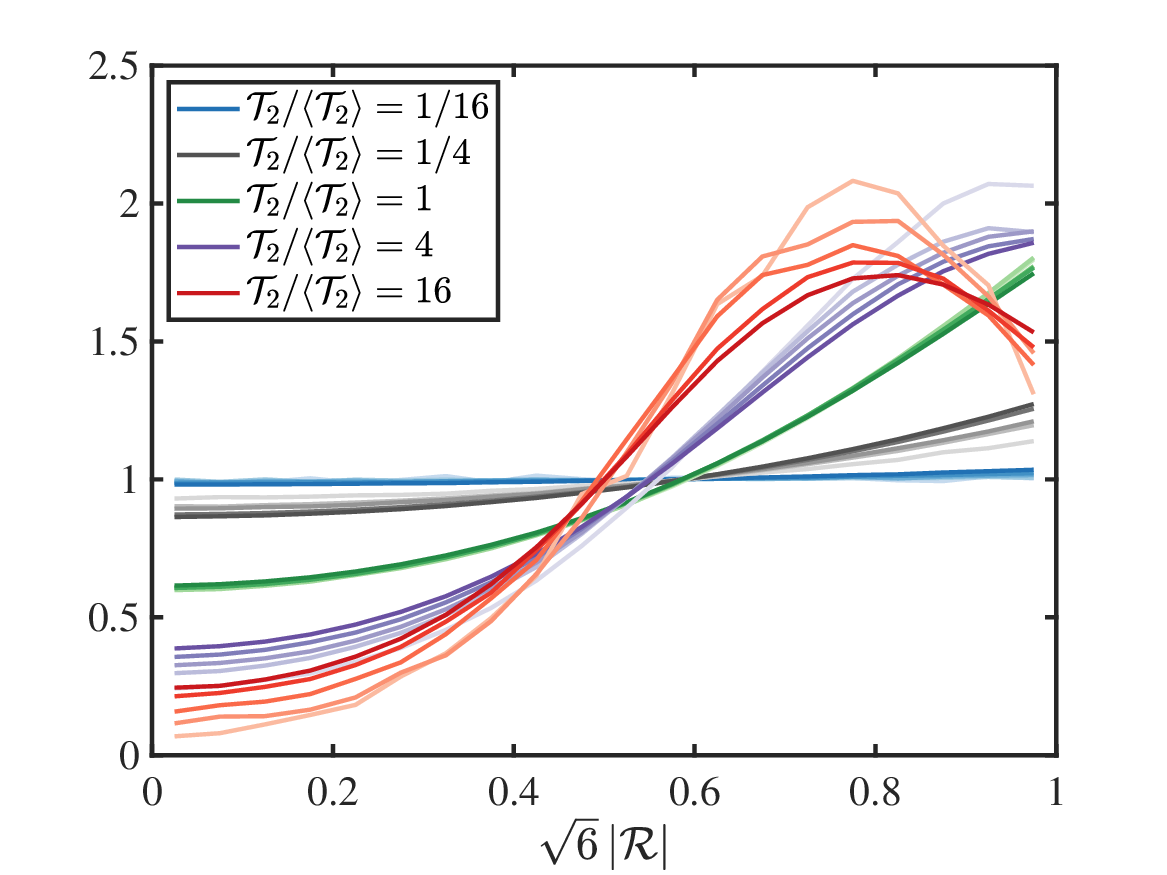}
}
\caption{ The distribution of (a) $\RR$ and (b) $|\RR|$, conditioned on
several values of $\TT_2/\langle \TT_2 \rangle$, for the runs described in
Appendix~\ref{appendixC} (see Table~\ref{tab:dns}), the convention chosen being such 
that the color
becomes brigher when $R_\lambda$ increases.
At very small values of $\TT_2/\langle \TT_2 \rangle$,
the distribution of $\RR$ is almost uniform, but increasingly deviates 
from a uniform distribution when
$\TT_2 /\langle \TT_2 \rangle$ becomes larger.
The values of $\TT_2 /\langle \TT_2 \rangle $ indicated on the legend
were obtained by averaging over all values in the interval $1/2^{4} $
and $2^{4}$ around the nominal value.
}
\label{fig:cond_R}
\end{center}
\end{figure}

In fact, Fig.~\ref{fig:cond_R}a shows the probability of $\sqrt{6} \RR$,
conditioned on a few values of $\TT_2/\langle \TT_2 \rangle$. At small values
of $\TT_2/\langle \TT_2 \rangle$, for $\TT_2/\langle \TT_2 \rangle = 1/16$,
the distribution of $\RR$ appears as very close to uniform. However, as
$\TT_2/\langle \TT_2 \rangle$ increases, the distribution gets increasingly
shifted to negative values of $\RR$. Fig.~\ref{fig:cond_R}b shows the
probability of $\sqrt{6}|\RR|$, which becomes increasingly peaked at values of
$|\RR| > 0.5$, therefore signaling that the value
$\langle \RR^2 | \TT_2 \rangle > 1/18$. This points to a value of
$\langle \TT_3^2 \TT_2^p \rangle/\langle \TT_2^{p+3} \rangle > 1/18$, therefore
implying a deviation from the formulae derived in~\cite{boschung:15}.

The distribution of values of $\RR$ when
$\TT_2/\langle \TT_2 \rangle \to 0$ is therefore very simple. As we will show in
the appendix~\ref{appendix:G}, the distribution of $\RR$ is indeed uniform
when taking a Gaussian ensemble of matrices. This remark shows that one
has to be careful about assumptions in the spirit of a Gaussian: these are
(empirically) valid only when $\TT_2 /\langle \TT_2 \rangle \ll 1$.

\begin{figure}[t]
\begin{center}
\includegraphics[scale=0.36]{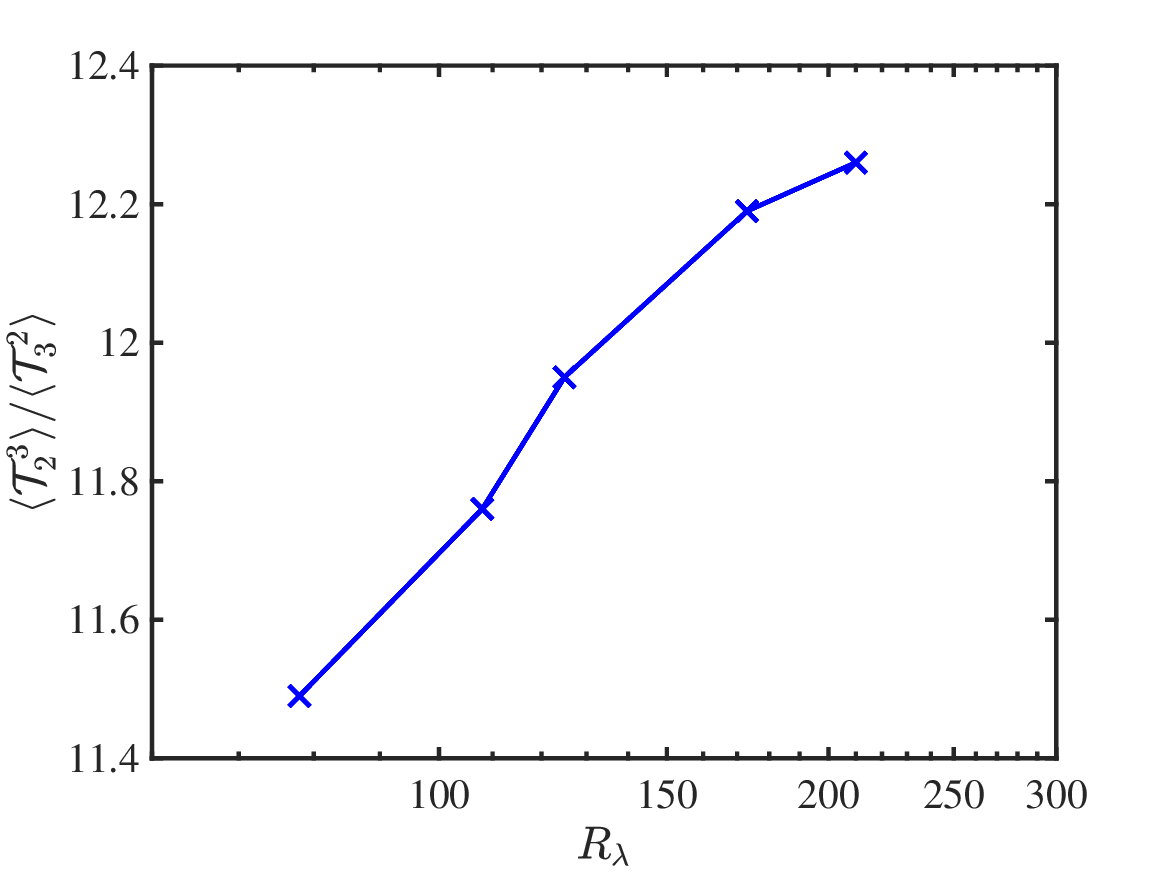}
\caption{ The ratio 
$\langle \TT_2^3 \rangle / \langle \TT_3^2 \rangle$
at several values of $R_\lambda$, corresponding to the runs listed in
Appendix~\ref{appendixC} (see Table~\ref{tab:dns}).
}
\label{fig:ratio}
\end{center}
\end{figure}

We determined numerically the statistical properties of the distribution of 
$\mathcal{R}$. 
Details about the numerical methods are provided in Appendix
~\ref{appendixC}. 
The excesss of probability for values of $\TT_2 / \langle \TT_2 \rangle \gtrsim 1$ implies that the ratio
$\langle \TT_3^2 \rangle / \langle \TT_2^3 \rangle > 1/18$.
Numerically, we find values closer to
$\langle \TT_2^3 \rangle / \langle \TT_3^2 \rangle \approx 12$,
plausibly increasing with $R_\lambda$, see Fig.~\ref{fig:ratio}. 
We have also computed the ratios 
$\langle \TT_2^{p+3} \rangle /\langle \TT_2^p \TT_3^2 \rangle$ for $p = 1$ 
and $2$, and found results comparable to those found in Fig.~\ref{fig:ratio}.
Our data also suggests that the ratio $\langle \TT_2^6 \rangle / \langle \TT_3^4 \rangle \sim 105$, significantly smaller than the value of $180$ obtained
in the Gaussian case. Although the values given here are indicative, in view
of the limitations of the DNS database, they clearly show that deviations from 
the simplifying assumptions leading to the results derived in~\cite{boschung:15}
are significant.

\subsection{Implications for terms of the form $\langle \TT_2^{n-3p} \TT_3^{2p} \rangle$ }
\label{subsec:T2^n-3p_T3^2p}
 
The analysis carried above leads to an explicit prediction for terms, appearing 
in the expression of moments of $S_{11}$ of order $n \ge 6$, of the form 
$\langle \TT_2^{n - 3p}  \TT_3^{2p} \rangle$ for $3 p \le n$. Assuming that the 
$\mathcal{R}$ is uniformly distributed in the interval 
$-1/\sqrt{6} \le \mathcal{R} \le 1/\sqrt{6}$, it is a simple matter to show
that:
\begin{align}
\frac{\langle \TT_2^{n-3p} \TT_3^{2p} \rangle}{\langle \TT_2^{n} \rangle } = \int_{-1/\sqrt{6}}^{1/\sqrt{6}}  \frac{\sqrt{6}}{2} \  u^{2p} \ du = \frac{1}{(2p+1) 6^{p} } \, ,\label{eq:mom_p}
\end{align}
which reduces to $1/18$ when $p = 1$. 
For $n = 6$, $7$ and $8$, the ratios $\langle \TT_2^{n-6} \TT_3^{4} \rangle$,
appearing in Eqs.~\eqref{eq:12_mom}, \eqref{eq:14_mom} and \eqref{eq:16_mom},
and $  \langle \TT_2^n \rangle$ are equal to $1/180$.
We notice that, if one takes into substitutes the expressions 
$\langle \TT_3^2 \TT_2^{n-3} \rangle /\langle \TT_2^n \rangle = 1/18 $
and
$\langle \TT_3^4 \TT_2^{n-6} \rangle /\langle \TT_2^n \rangle = 1/180 $
in Eqs.~\eqref{eq:12_mom}, one recovers exactly the expression for the moment 
of order $12$ given by Eq.~\eqref{eq:mom_ps}.

\section{Discussion and conclusions}

 In this note, we established the relation between the moments of the even 
moments of the longitudinal velocity gradient, $\partial_x u = S_{11}$, and
the invariants of the tensor $\SSs$, namely $\TT_2$ and $\TT_3$. 
Whereas the moments of order $2$ and $4$ are simply related to the mean
and quadratic moments of $\TT_2$, the expression of moments of
$S_{11}$ of order $n \ge 6$ involves terms such as $\langle \TT_3^2 \TT_2^{n - 6} \rangle$, and for moments of order $n \ge 12$, terms of order 
$\langle \TT_2^{n-12} \TT_3^4 \rangle$ etc. 
In this sense, our work contradicts the recent derivation by 
Boschung~\cite{boschung:15} of a simple relation between 
$\langle S_{11}^{2n} \rangle$ and $\langle \TT_2^n \rangle$, without any role 
of $\TT_3^{2p}$.
The difference between those simplified formulae
comes from the implicit assumption that the ratio 
$\mathcal{R} = \frac{{\rm tr} (\SSs^3)}{{\rm tr} (\SSs^2)^{3/2} }$ is
uniformly distributed in the interval $[ -1/\sqrt{6}, 1/\sqrt{6}]$, as it is
the case when considering an Gaussian ensemble of random matrices,
as discussed in Appendix~\ref{appendix:G}. Indeed,
the derivation of~\cite{boschung:15} explicitly uses the generating function,
which is appropriate for a Gaussian ensemble.
It is known, however, that the bias
in the distribution of the intermediate eigenvalue of $\SSs$, which
tends to be positive, is an essential
aspect of velocity gradient amplification (vortex stretching),
an essential aspect of intermittency~\cite{betchov:56}, which in turn implies
$\langle \omega_i \omega_j S_{ij} \rangle = - \frac{4}{3} \langle \TT_3 \rangle$~\cite{betchov:56}.
This relation points to a nontrivial distribution of $\mathcal{R}$, and indeed,
to a difference with the relation derived in~\cite{boschung:15}.
As a consequence, the effect discussed in these notes establishes that this 
bias in the
distribution of the intermediate eigenvalue of strain affects the moments
of $S_{11}$ of order $n \ge 6$.

The slight deviation between
$\rho_3 \equiv \langle \TT_3^2 \rangle/\langle \TT_2^3 \rangle$
and $1/18$ generates a small error when estimating the $6^{th}$ moment
of $S_{11}$ using Eq.~\eqref{eq:mom_ps}. Specifically,
rewriting, in the case of the $6^{th}$ moment:
\begin{align}
\langle ( S_{11})^6 \rangle
=  \langle \TT_2^3 \rangle \left( \frac{40}{3003} + \frac{128}{9009} \frac{\langle \TT_3^2 \rangle}{\langle \TT_2^3 \rangle}  \right)
=  \langle \TT_2^3 \rangle \left( \frac{8}{567}
+ \frac{128}{9009} \left[ \frac{\langle \TT_3^2 \rangle}{\langle \TT_2^3 \rangle} - \frac{1}{18}   \right] \right) \, ,
\label{eq:re_6_mom}
\end{align}
which provides an explicit estimate of error in the expression 
of~\cite{boschung:15}.
Namely, taking a value of
$\langle \TT_3^2 \rangle/\langle \TT_2^3 \rangle = 1/12$,
as suggested by Fig.~\ref{fig:ratio}, we find that the correction term,
$128/9009 ( 1/12-1/18)$, is less than $6 \%$ compared to the dominant
term, $8/567$. Therefore, a direct measurement of the corrections to
Eqs.~(\ref{eq:mom_ps}) require a very accurate
determination of the various moments.

Here we discussed the relation between the moments of 
dissipation, $\langle \epsilon^n \rangle$, and of its surrogate, $\langle \epsilon_s^n \rangle$. 
An equivalent question is to derive an exact relation between the PDFs of 
$\epsilon$ and $\epsilon_s$.
The results presented in this work show that an \emph{exact and closed} relation cannot be obtained. 
On the other hand, the very small numerical difference between the results predicted under the Gaussian assumption~\cite{boschung:15} suggests that
an approximate relation might be derived. In fact, for a random vector in an isotropic field, there exists an exact relation between the PDFs of its magnitude and its components, as has been shown for accelerations \cite{mordant:2004} and vorticities \cite{gotoh:25} in HIT. Although this exact relation cannot be extended to random tensors,
T. Gotoh and P. K. Yeung proposed a relation between the PDFs of $\epsilon$ and $\epsilon_s$, which 
seems to capture very accurately the numerical results \cite{gotoh:26}.

In this work, we did not explicitly address the odd moments of $S_{11}$, 
which in fact can be obtained by the very same approaches~\cite{wu:26}.  
Here, we merely note that for HIT in a statistically steady state, 
the third moment $\langle \TT_3 \rangle$ is related to vorticity 
amplification (vortex streching~\cite{betchov:56}). Expressing that 
enstrophy dissipation balances enstrophy production leads to the following
relation between $\langle \TT_3 \rangle$ and the second derivative of velocity:
\begin{equation}
\langle \TT_3 \rangle = \langle {\rm tr}( \SSs^n )  \rangle = \frac{105}{8} \langle \left(\frac{\partial u_1}{\partial x_1} \right)^3 \rangle = - \frac{105}{4} \nu \langle \left(\frac{\partial^2 u_1}{\partial x^2_1} \right)^2 \rangle \ ,
\label{eq:vel_hess}
\end{equation}
see, e.g., Eq.~(5.37) in Davidson~\cite{Davidson:2015}. Alternatively,
Eq.~\eqref{eq:vel_hess} can be obtained by combining the Karman-Lin equation 
for the energy balance in wavenumber space and equations (12.79) and 
(12.144')\footnote{The printed version of Eq.~(12.144') in Monin \& Yaglom contains a typo. The coefficient $-\frac{35}{4}$ should be $-
\frac{35}{2}$} in Monin \& Yaglom~\cite{MoninYaglom:v2}.
Equation~\eqref{eq:vel_hess} relates the third moment 
$\langle \TT_3 \rangle$ of the velocity derivative to its spatial 
variations, as a consequence of turbulence dynamics. The
highly nonlinear character of this dynamics makes it unlikely that 
the statistics of the velocity gradients can be captured by a Gaussian 
approximation.

\acknowledgements

We are grateful to T. Gotoh for discussions, which inspired this work, 
with thank to him and P. K. Yeung for communicating their preliminary results to us~\cite{gotoh:26}. 
As we were preparing the manuscript, we became aware of the work of T. Wu and M. Wilczek, whom we thank for sharing 
their insight~\cite{wu:26}. 
We are also grateful to D. Buaria and K. Kozlowski for discussions. 
PY and HX acknowledge support from the National Natural Science Foundation
of China (NSFC) under grants 12572254, 12202452 and 12588201. PY also
acknowledges support from the Fundamental Research Funds for Central University (D5000240110). 
AP was supported by the French Agence Nationale de la Recherche under 
Contract No. ANR-20-CE30-0035 (project TILT), and in part by NSF Grant No.
PHY-2309135 to the Kavli Institute for Theoretical Physics (KITP). 
AP is also thankful to the 
Max Planck Institute for Dynamics and Self-organisation (G\"ottingen,
Germany) for continuous support, and and to 111 Project that funded 
his collaborations with Tsinghua University and other institutions in China.

\appendix

\section{Distribution of $\RR$ for Gaussian matrices}
\label{appendix:G}

The matrix $\SSs$ in a turbulent flow can be considered as random. Although
the evidence accumulated over the years points to strong deviations from
the distribution of the velocity gradients from a Gaussian distribution,
it is instructive to consider the distribution of eigenvalues of an
ensemble of symmetric, traceless Gaussian matrices.

 Such an ensemble is fully defined by the
variances of the matrix elements, which can be derived by expressing
$\langle S_{ij} S_{kl} \rangle = A \delta_{ij} \delta_{kl}
+ B ( \delta_{ik} \delta_{jl} + \delta_{il} \delta_{jk} )$, and by expressing
incompressiblility as well as the condition that $\langle {\rm tr} (\SSs^2) \rangle = \epsilon/(2 \nu)$. This leads to (we do not imply here summation over
repeated greek indices in the following expression):
\begin{align}
 \langle S_{\alpha \alpha}^2 \rangle = \frac{1}{15} \sigma^2 ~ , ~~ \langle S_{\alpha \alpha} S_{\beta \beta} \rangle = - \frac{1}{30} \sigma^2~ , ~~~ \langle S_{\alpha \beta} S_{\alpha \beta} \rangle = \frac{1}{20} \sigma^2  \, ,
\label{eq:var}
\end{align}
where $\sigma^2 = \epsilon/\nu$. Note that Eq.~\eqref{eq:var} implies that
$\langle (S_{11} + S_{22} + S_{33})^2 \rangle = 3 \langle S_{11}^2 \rangle + 6 \langle S_{11} S_{22} \rangle = 0$.

Instead of parametrizing the matrix by its components $S_{ij}$ it is more
appropriate to return to the diagonalization introduce above:
$\SSs = \Rot \DD \Rot^{T}$, where $\Rot$ is a rotation matrix, and
$\DD$ is a diagonal matrix with the three eigenvalues,
$\DD = {\rm diag} (\lambda_1 , \lambda_2 , \lambda_3)$. The transformation
between the parametrization in terms of the $5$ distinct components,
$S_{11}$, $S_{22}$ and $S_{ij}$ for $i < j$ on the one hand, and the
parametrization given by the 2 independent eigenvalues, $\lambda_1$ and
$\lambda_2$ ($\lambda_3 = - (\lambda_1 + \lambda_2)$) , and the 3-dimensionsal
rotation group of rotation matrix, is~\cite{deift:09,Kozlowski:25}:
\begin{align}
 \frac{ \partial (S_{ij} ) }{\partial{ ( \lambda_1, \lambda_2, \Rot )}} = \prod_{1 \le i<j \le 3} | \lambda_i - \lambda_j|  \times F( \Rot ) \, ,
\label{eq:Jacobian}
\end{align}
where $F( \Rot )$ is the probability density on the rotation group.

An alternative parametrization of the matrix $\SSs$ can be done in terms of
the two invariants, $\TT_2/2$ and $\TT_3/3$. The relation between
$( \TT_2/2 ,  \TT_3/3)$ and $( \lambda_1, \lambda_2)$ is simply:
\begin{align}
\frac{\TT_2}{2} = \lambda_1^2 + \lambda_2^2 + \lambda_1 \lambda_2 ~~~~ {\rm and} ~~~~ \frac{\TT_3}{3} = - ( \lambda_1^2 \lambda_2 + \lambda_1 \lambda_2^2) ,.
\label{eq:lambda_T}
\end{align}
Elementary algebra shows that the Jacobian of the transformation
$(\lambda_1 , \lambda_2 ) \to (\TT_2/2, \TT_3/3)$ is:
\begin{eqnarray}
\frac{ D ( \TT_2/2 , \TT_3/3 )}{D (\lambda_1, \lambda_2)} & = & | {\rm det}
\begin{pmatrix}
2 \lambda_1 + \lambda_2 & 2 \lambda_1 \lambda_2 + \lambda_2^2 \\
2 \lambda_2 + \lambda_1 & 2 \lambda_2 \lambda_1 + \lambda_1^2
\end{pmatrix} |  \nonumber \\
& = & | (2 \lambda_1 + \lambda_2) ( 2 \lambda_2 + \lambda_1) (\lambda_2 - \lambda_1) | = | (\lambda_3 - \lambda_1) (\lambda_3 - \lambda_2) (\lambda_2 - \lambda_1) |  \nonumber \\
& = & \prod_{1 \le i < j \le 3} | \lambda_i - \lambda_j | \, .
\label{eq:jacob_trsf}
\end{eqnarray}

Therefore, the parametrization of the matrix $\SSs$ in terms of
the two invariants, $\TT_2/2$ and $\TT_3/3$, and the rotation matrix $\Rot$,
gives right to the Jacobian:
\begin{align}
 \frac{ \partial (S_{ij} ) }{\partial{ ( \TT_2/2, \TT_3/3, \Rot )}} = F( \Rot ) \, .
\end{align}
The standard distribution of $\SSs$ is given by an exponential probability
weight: $\propto \exp( - \TT_2 )$. In which case, when conditioning the
properties of
$\TT_3/3$ (in particular the distributions of eigenvalues) on $\TT_2/2$,
one finds that the distribution of $\RR$ is uniform,
as also found in the case of the matrix $\SSs$ when $\TT_2/\langle \TT_2 \rangle \ll 1$, as shown in Fig.~\ref{fig:cond_R}.

\section{Algorithmic Determination of the Coefficients in Eq.~\eqref{eq:mom_general}}
\label{app:mathematica}

\begin{lstlisting}[language=Mathematica]
(* =========================*)
(*0. Clear symbols*)
(* =========================*)
ClearAll[n, \[Lambda]1, \[Lambda]2, \[Lambda]3, c1, c2, c3, df,
  angleAvgK, F, T2expr, T3expr, G, solveA];

(* =========================*)
(*1. Double factorial and angular average formula*)
(* =========================*)

(*double factorial, with (-1)!!=1*)
df[m_Integer] := If[m <= 0, 1, Times @@ Range[m, 1, -2]];

(* <c1^(2a) c2^(2b) c3^(2c)> *)
angleAvgK[a_Integer, b_Integer, c_Integer, n_Integer] :=
  df[2 a - 1] df[2 b - 1] df[2 c - 1] / df[4 n + 1];

(* =========================*)
(*2. Compute F[n] = <S11^(2n)>*)
(* =========================*)

(* Use: S11 = \[Lambda]1 c1^2 + \[Lambda]2 c2^2 + \[Lambda]3 c3^2,
   k1 + k2 + k3 = 2n, sum over k1,k2 with k3=2n-k1-k2 *)
F[n_Integer?Positive] :=
  Module[{k1, k2, k3, expr},
   expr = Sum[
     k3 = 2 n - k1 - k2;
     If[k3 < 0, 0,
      Multinomial[k1, k2, k3] *
        angleAvgK[k1, k2, k3, n] *
        \[Lambda]1^k1 * \[Lambda]2^k2 * \[Lambda]3^k3
      ],
     {k1, 0, 2 n}, {k2, 0, 2 n - k1}
     ];

   (* incompressibility constraint: \[Lambda]3 = -\[Lambda]1 - \[Lambda]2 *)
   Expand[expr /. \[Lambda]3 -> -\[Lambda]1 - \[Lambda]2]
   ];

(* =========================*)
(*3. T2, T3 and basis functions G[n,p]*)
(* =========================*)

(* For a 3x3 incompressible symmetric traceless matrix:
   T2 = tr(S^2) = 2(\[Lambda]1^2 + \[Lambda]2^2 + \[Lambda]1 \[Lambda]2)
   T3 = tr(S^3) = 3 \[Lambda]1 \[Lambda]2 \[Lambda]3
               = -3 \[Lambda]1 \[Lambda]2(\[Lambda]1 + \[Lambda]2) *)
T2expr = 2 (\[Lambda]1^2 + \[Lambda]2^2 + \[Lambda]1 \[Lambda]2);
T3expr = -3 \[Lambda]1 \[Lambda]2 (\[Lambda]1 + \[Lambda]2);

(* G[n,p] = T2^(n-3p) T3^(2p) *)
G[n_Integer?Positive, p_Integer?NonNegative] :=
  Expand[T2expr^(n - 3 p) T3expr^(2 p)];

(* =========================*)
(*4. Solve A_p using numerical sampling*)
(* =========================*)

solveA[n_Integer?Positive] :=
 Module[{maxp, num, pts, Fvals, Mmat, Avec},
  maxp = Floor[n/3];
  num = maxp + 1;  (* number of sampling points *)

  (* choose non-special (\[Lambda]1,\[Lambda]2) values *)
  pts = Table[{i, i + 1}, {i, 1, num}];

  (* compute F[n](\[Lambda]1, \[Lambda]2) at sampling points *)
  Fvals = Table[
    F[n] /. {\[Lambda]1 -> pts[[k, 1]], \[Lambda]2 -> pts[[k, 2]]},
    {k, 1, num}
    ];

  (* build matrix M_{k,p} = G[n,p](\[Lambda]1_k,\[Lambda]2_k) *)
  Mmat = Table[
    G[n, p] /. {\[Lambda]1 -> pts[[k, 1]], \[Lambda]2 -> pts[[k, 2]]},
    {k, 1, num}, {p, 0, maxp}
    ];

  (* solve linear system M A = F *)
  Avec = LinearSolve[Mmat, Fvals];

  (* return A_p -> rational numbers *)
  Table[Ap[p] -> Rationalize[Avec[[p + 1]], 0], {p, 0, maxp}]
  ];
\end{lstlisting}

\section{Numerical data}
\label{appendixC}. 

To clarify the role of the distribution of $\mathcal{R}$, defined by Eq.~\eqref{eq:def_rho}, we systematically determined the distribution of $\mathcal{R}$
from DNS at moderate resolution. The data was obtained using a standard
pseudo-spectral code, in a triply periodic domain, of size $(2\pi)^3$, with $N$ 
grid points, as shown in Table~\ref{tab:dns}. The code is dealiased using
the algorithm proposed by~\cite{Patterson:71}. 
The quality of the resolution can be judged by the
product $k_{max} \eta$, given in Table~\ref{tab:dns}. The data was generated
by analyzing $N_s$ instantaneous full DNS fields  ($N_s \sim 20$), saved during the runs, covering a time $T_{run}$
at least equal to $\gtrsim 15 T_E$, where $T_E$ is the eddy turnover time.
A more precise description of the code can be found e.g. in~\cite{Pumir94}.

\begin{table}[h]
\centering
 \begin{tabular}{c|ccccc}
\hline
 $\re$  &  $N $ & $k_{max} \eta$ &
$T_{run}/T_E$ & $N_s$  \\
\hline
$78$ & 128 & 1.6 & 46.4 & 17   \\
$108$ & 192 & 1.6 &  19.9 & 18 \\
$125$ & 256 & 1.6 &  23.2 & 17 \\
$173$ & 384 & 1.6 &  22.6 & 30 \\
$210$ & 512 & 1.6 &  16.5 & 38 \\
\hline
\end{tabular}
\caption{Main characteristics of the runs.
}
\label{tab:dns}
\end{table}

\bibliography{references}

\end{document}